\documentclass[aps,prc,10pt,notitlepage,nofootinbib,superscriptaddress]{revtex4-1}

\usepackage[utf8]{inputenc}
\usepackage{amsmath}
\usepackage{amssymb}
\usepackage{amsfonts}
\usepackage{newtxtext,newtxmath}
\usepackage{bm}
\usepackage{graphicx}
\usepackage[usenames,dvipsnames]{xcolor}
\usepackage{color}
\usepackage[colorlinks=true,linkcolor=blue,urlcolor=blue,citecolor=blue]{hyperref}

\usepackage{slashed}
\usepackage[english]{babel}
\usepackage{dcolumn}
\usepackage{blindtext}
\usepackage{epsfig}
\usepackage{pifont}
\usepackage{dsfont}
\usepackage{cancel}
\usepackage{bigints}
\usepackage{accents}
\usepackage{soul}
\usepackage{multirow}
\usepackage{simpler-wick}

        \usepackage{setspace}

\newcommand{\nn}{\nonumber}
\newcommand{\ket}[1]{\left|#1\right\rangle}

\newcommand{\FB}[1]{\left(#1\right)}

\newcommand{\SB}[1]{\left\{#1\right\}}
\newcommand{\TB}[1]{\left[#1\right]}

\newcommand{\sign}[1]{\text{sign}\left(#1\right)}

\newcommand{\del}{\partial}

\begin{document}
\title{Finite size effect on the thermodynamics of a hot and magnetized hadron resonance gas}

\author{Debasis Atta}
\email{debasisa906@gmail.com}
\thanks{Corresponding Author}
\affiliation{Government General Degree College Kharagpur-II, Paschim Medinipur - 721149, West Bengal, India}

\author{Nilanjan Chaudhuri}
\email{sovon.nilanjan@gmail.com, n.chaudhri@vecc.gov.in}
\affiliation{Variable Energy Cyclotron Centre, 1/AF Bidhannagar, Kolkata - 700064, India}
\affiliation{Homi Bhabha National Institute, Training School Complex, Anushaktinagar, Mumbai - 400085, India}

\author{Snigdha Ghosh}
\email{snigdha.physics@gmail.com, snigdha@ggdckharagpur2.ac.in}
\thanks{Corresponding Author}
\affiliation{Government General Degree College Kharagpur-II, Paschim Medinipur - 721149, West Bengal, India}

\begin{abstract}
The thermodynamic properties of a non-interacting ideal Hadron Resonance Gas (HRG) of finite volume have been studied in the presence of an external magnetic field. The inclusion of background magnetic field in the calculation of thermodynamic potential is done by the modification of the dispersion relations of the charged hadrons in terms of Landau quantization. The generalized Matsubara prescription has been employed to take into account the finite size effects in which a periodic (anti-periodic) boundary conditions is considered for the mesons (baryons). We find significant effects of the magnetic field as well as system size on the temperature dependence of energy density, longitudinal and transverse pressure especially in low temperature regions. The HRG is found to exhibit diamagnetism (paramagnetism) in the low (high) temperature region whereas the finite size effect is seen to strengthen the diamagnetic behavior of the medium.
\end{abstract}

\maketitle
%
\section{INTRODUCTION}\label{sec.intro}

As an essential part of the Standard Model of particle physics, quantum chromodynamics (QCD) is the established theory of strong interaction between quarks and gluons. One of the most interesting predictions of QCD is that strongly interacting matter possess at least two distinct phases: at low temperature and/or baryon density it consists of colourless hadrons, while at high values of temperature and/or baryon density the fundamental degrees of freedom are deconfined, asymptotically free coloured quarks and gluons, commonly known as quark-gluon plasma (QGP).  On the theoretical perspective, a first principle QCD based calculation of the thermodynamics of `strongly' interacting matter is highly restricted due to large QCD coupling ($\alpha_s$) especially in the low temperature region. The numerical Lattice QCD (LQCD) simulations provide suitable framework to deal with this non-perturbative aspect of QCD in the low baryon density region of the QCD phase diagram which indicates that a smooth crossover exists from the low temperature hadron gas phase to a quark-gluon plasma, at a (pseudo)-critical temperature $ T_c \approx 155  $~MeV~\cite{Borsanyi:2010cj,Borsanyi:2012cr,Borsanyi:2013bia,Bazavov:2014pvz}. However, the present LQCD simulations  are plagued with the so-called “sign problem” in case finite chemical potential~\cite{Loh:1990zz,Chandrasekharan:1999cm}. The application of QCD to Big Bang cosmology demonstrates that as the early universe expanded and cooled, it must underwent a transition from the QGP to a confined hadronic phase, roughly $ 10^{-5}  $ seconds after Big Bang~\cite{Witten:1984rs,Kajantie:1986hq,Fuller:1987ue}. On the other hand, in Heavy-Ion Collision (HIC) experiments conducted at the Relativistic Heavy Ion Collider (RHIC) and the Large Hadron Collider (LHC), the same transition is expected to occur, which provides a remarkable opportunity to study the phase structure or the bulk thermodynamic properties of QCD over a wide range of temperature and baryon density~\cite{STAR:2005gfr,PHENIX:2004vcz,PHOBOS:2004zne,BRAHMS:2004adc,Braun-Munzinger:2015hba}. One of the major objectives of these HIC experiments is to determine how the thermodynamic observables, such as pressure, energy density and so on, change as the system evolves through the region separating the hadronic and QGP phase. In presence of thermodynamics equilibrium, intermediate relations between these observables, constitute the equation of state (EoS) of the system. The functional relationship between EoS and control parameters, such as temperature and chemical potential, is one of the necessary inputs of the relativistic hydrodynamic simulations of the hot and dense matter created in HIC experiments~\cite{Teaney:2000cw,Romatschke:2009im,Kolb:2003dz,Jacobs:2004qv}.

 Moreover, the study of the influence of  a uniform background magnetic field on various microscopic as well as bulk properties of QCD matter at extreme condition of high temperature and/or baryon density, has been propelled to the forefront in  recent years (see Refs.~\cite{Kharzeev:2013jha,Miransky:2015ava,Andersen:2014xxa} for review).  Recent studies~\cite{Kharzeev:2007jp,Skokov:2009qp} indicate that, in a non-central or asymmetric HIC experiment, very strong magnetic fields of the order $ \sim 10^{18} $ Gauss or larger might have been produced by the spectators due to the collision geometry. Although, these field are time dependent, the presence of large electrical conductivity of the hot and dense magnetized medium results in  substantial delay in the decay process of the generated magnetic fields~\cite{Tuchin:2013apa,Tuchin:2015oka,Tuchin:2013ie,Gursoy:2014aka}.  Since, the strength of the magnetic field reaches up to the typical QCD energy scale ($eB\sim \Lambda_\text{QCD}^2$), the EoS as well as bulk properties of the QCD matter could be significantly altered~\cite{Kharzeev:2013jha}.  Apart from this, strong magnetic fields are expected to be present in several other physical objects. For example, in the interior of certain astrophysical objects called magnetars~\cite{Duncan:1992hi,Thompson:1993hn}, magnetic field $\sim 10^{15}$ Gauss can be present. Furthermore, it is conjectured using Cosmological model calculations that, in the early universe during the electroweak phase transition, magnetic fields as high as $ \sim 10^{23}$ Gauss~\cite{Vachaspati:1991nm,Campanelli:2013mea} might have been created.  In each of these cases, the interplay between dynamics of QCD matter and background magnetic field results in plethora of novel and exciting phenomena; for example, the Chiral Magnetic Effect (CME)~\cite{Fukushima:2008xe,Kharzeev:2007jp,Kharzeev:2009pj}, Magnetic Catalysis (MC)~\cite{Shovkovy:2012zn,Gusynin:1994re,Gusynin:1995nb,Gusynin:1999pq}, Inverse Magnetic Catalysis (IMC)~\cite{Preis:2010cq,Preis:2012fh}, Chiral Vortical Effect (CVE), vacuum superconductivity and superfluidity~\cite{Chernodub:2011gs,Chernodub:2011mc} {\it etc}.

Consequently, a precise theoretical knowledge about the modifications of the EoS as a function of state parameters, such as, temperature, chemical potential and magnetic field, is desirable  among the large number researchers in this domain of physics. A huge part of our current understanding about the dependence of EoS on background magnetic field at vanishing chemical potential comes from LQCD studies~\cite{Bali:2014kia,Endrodi:2014vza,Endrodi:2014lja,Endrodi:2015oba,Endrodi:2015vzl,Tawfik:2016gye}. However, a simpler alternative is provided by effective models for the study of the strongly interacting matter in the non-perturbative domain. The hadron resonance gas (HRG) model~\cite{Hagedorn:1980kb,Rischke:1991ke,Cleymans:1992jz,Vovchenko:2014pka,Andronic:2012ut,Endrodi:2013cs} is a statistical thermal model which is capable of studying the QCD thermodynamics at finite temperature, baryon density and external magnetic field. Moreover, in the absence of the external magnetic field,  it has been found that, the results from HRG model agrees well with the thermodynamic observables derived from LQCD simulations even up to temperatures just below the transition region, at both zero~\cite{Borsanyi:2010cj,Karsch:2003vd,Huovinen:2009yb} and non-zero~\cite{Borsanyi:2012cr,Karsch:2003zq,Tawfik:2004sw,Allton:2005gk} baryon chemical potential. This simple model has been extensively used in many physical situations  at zero magnetic field, for example, description of different hadron yields in HIC experiments from AGS up to RHIC energies~\cite{Cleymans:1999st,BraunMunzinger:1994xr,BraunMunzinger:1999qy,BraunMunzinger:2001ip,Andronic:2005yp,Andronic:2008gu,Becattini:2005xt,Becattini:2012xb,Chatterjee:2013yga}, conserved charge fluctuations~\cite{Nahrgang:2014fza,HotQCD:2012fhj,Begun:2006jf}, transport coefficients for hadronic matter~\cite{Prakash:1993bt,Gorenstein:2007mw,Noronha-Hostler:2008kkf,Puglisi:2014pda,Kadam:2014cua,Greif:2016skc,Ghosh:2019fpx}. The modifications of thermodynamic as well as transport  properties of QCD matter in the presence of the magnetic field has been examined in several papers~\cite{Endrodi:2013cs,Bhattacharyya:2015pra,Kadam:2014xka,Mohapatra:2017zrj,Das:2019pqd,Dash:2020vxk}.

In all the works mentioned above, the finite volume effects on the phase structure of strongly interacting matter have not been considered. However, the results from the HIC experiments, suggests that the created fireball has finite spatial volume. For example, in~\cite{Graef:2012sh}, examining Pb-Pb collisions using UrQMD model, it was shown that the freeze out volume can be $ \sim 50 $ to $ 250  $ fm$ ^3 $ depending on different energies and centralities. Since these are freeze out volume, at early phases of the evolution the system size is expected to be smaller (few fm$^3$)~\cite{STAR:2005gfr,CERES:2002rfr}. So, the quantitative measurements of the observables related to both QGP and hadron phase can be significantly modified owing to the finite size of the fireball. Thus the influence of the finite volume of the system on the thermodynamic observables and phase structure of the strongly interacting matter has been thoroughly analysed using various approaches in the literature for both zero and finite values of the background magnetic field. Such as, Dyson-Schwinger equations of QCD~\cite{Luecker:2009bs,Li:2017zny,Shi:2018swj,Shi:2020uyb}, renormalization group aided quark-meson model~\cite{Braun:2004yk,Braun:2005fj,Klein:2017shl,Wan:2020vaj}, the Nambu–Jona-Lasinio (NJL) model and its extensions~\cite{Palhares:2009tf,Wang:2018qyq,Pan:2016ecs,Bhattacharyya:2012rp,Ferrer:1999gs,Abreu:2019czp}  and others~\cite{Wu:2017xaz,Abreu:2017lxf,Damgaard:2008zs,Abreu:2013nca,Abreu:2014ofa,Abreu:2015jya,Grunfeld:2017dfu}. A detailed study of the finite volume effect in ambit of HRG model has been made in~\cite{Bhattacharyya:2015zka} and a significant difference is observed in the fluctuations for electric charge. However,  we have not come across any previous calculations regarding modifications of thermodynamic properties of a magnetized HRG model due to consideration of finite size system.

In this work, we aim to study the thermodynamics of an ideal HRG of finite size in the presence of external magnetic field. For this, the thermodynamic potential will be modified to take into account the Landau quantization of the dispersion relation of charged hadrons due to background magnetic field. To consider the finite size effect, we will be using the generalized Matsubara prescription in which we will implement a periodic (anti-periodic) boundary conditions for the mesons (baryons).

The article is organized as follows. In Sec.~\ref{sec.formalism}, we discuss the formalism of calculating the HRG-thermodynamics at non-zero magnetic field for a finite sized system. Next in Sec.~\ref{sec.results}, we present the numerical results. Finally, we summarize and conclude in Sec.~\ref{sec.summary}.

\section{FORMALISM} \label{sec.formalism}
Let us start with the standard expression of the thermodynamic potential (density) $\Omega$ of an ideal HRG at zero-magnetic field in infinite volume as~\cite{Bhattacharyya:2013oya,Bhattacharyya:2015zka,Ghosh:2018nqi}
\begin{eqnarray}
\Omega = \Omega_\text{vac} -T \sum_{i\in\{\text{hadrons}\}}^{}g_ia_i\int\frac{d^3k}{(2\pi)^3}\ln\FB{1+a_ie^{-\beta\omega_k^i}}
\label{eq.Pot.0}
\end{eqnarray}
where $\Omega_\text{vac}$ is the vacuum contribution to the thermodynamic potential, 
$g_i = (2s_i+1)$ is the spin degeneracy of hadron $i$ having spin $s_i$, 
\begin{eqnarray}
a_i = -(-1)^{2s_i} = \begin{cases}
+1 ~~\text{if}~~ i\in\{\text{baryons}\}~~\text{(half-integer spin)} \\
-1 ~~\text{if}~~ i\in\{\text{mesons}\} ~~\text{(integer spin)},
\end{cases}
\end{eqnarray}
$\beta=1/T$ is the inverse temperature and $\omega_k^i = \sqrt{\vec{k}^2+m_i^2}$ is the single particle energy of hadron $i$ having mass $m_i$. 
We note that, in Eq.~\eqref{eq.Pot.0}, the explicit form of the quantity $\Omega_\text{vac}$ reads 
\begin{eqnarray}
	\Omega_\text{vac} = - \sum_{i\in\{\text{hadrons}\}}^{}g_ia_i\int\frac{d^3k}{(2\pi)^3} \omega_k^i 
	\label{eq.Pot.Vac}
\end{eqnarray}	
which is divergent and independent of temperature. It is worth mentioning that, in the calculation of physical quantities (like pressure, energy density, etc.), the contribution from this divergent vacuum term will be subtracted out owing to defining the normalized thermodynamic variables~\cite{Avancini:2020xqe,Chaudhuri:2020lga}.

From Eq.~\eqref{eq.Pot.0}, all the other thermodynamic quantities can be calculated. For example, the iostropic pressure is $P=-\Omega$, energy density ($\varepsilon$) is 
\begin{eqnarray}
\varepsilon = -T^2\frac{\del}{\del T}\FB{\frac{\Omega}{T}} = \varepsilon_\text{vac} +  \sum_{i\in\{\text{hadrons}\}}^{}g_i\int\frac{d^3k}{(2\pi)^3}\frac{\omega_k^i}{e^{\beta\omega_k^i}+a_i},
\label{e0}
\end{eqnarray}
where, $\varepsilon_\text{vac}$ is the contribution from $\Omega_\text{vac}$ and the entropy density is $s=(\varepsilon+P)/T$ etc.

Let us now consider an external magnetic field $\vec{B}=B\hat{z}$ in the positive-$\hat{z}$ direction. The single particle energies of charged hadrons will now be Landau quantized as
\begin{eqnarray}
\omega^i_{kls} = \sqrt{k_z^2+\{2l+1-2s~\sign{e_i}\}|e_i|B+m_i^2}~~~~\text{with}~~ s=-s_i,-s_i+1,\cdots,s_i ~~\text{and}~~ l=0,1,2, \cdots
\label{eq.disp.B}
\end{eqnarray} 
where, $e_i$ is the electronic charge of hadron $i$. It is to be noted that, in Eq.~\eqref{eq.disp.B}, $l$ is related to the orbital angular momentum quantum number and not the Landau level index in the usual sense though they are inter-connected. To obtain the expressions of the thermodynamic quantities at $B\ne0$, we make the following standard replacement of the phase-space integral as~\cite{Andersen:2014xxa}
\begin{eqnarray}
\int\frac{d^3k}{(2\pi)^3} f(\omega_k^i) \to \sum_{l=0}^{\infty}\sum_{s=-s_i}^{s_i} \frac{|e_i|B}{2\pi}\int_{-\infty}^{\infty}\frac{dk_z}{2\pi} f(\omega^i_{kls}).
\label{eq.phase.space}
\end{eqnarray} 
With the above replacement, the thermodynamic potential of ideal HRG of Eq.~\eqref{eq.Pot.0} in presence of external magnetic field modifies to
\begin{eqnarray}
\Omega_B = \Omega^B_\text{vac} + \Omega_{\!\!\!\begingroup
\renewcommand*{\arraystretch}{0.55}		
\begin{array}{c}
\text{\scriptsize neutral} \\ \text{ \scriptsize hadrons}
\end{array}
\endgroup}
 -T \sum_{i\in\SB{ \!\!\!
\begingroup
\renewcommand*{\arraystretch}{0.55}		
		\begin{array}{c}
	\text{\scriptsize charged} \\ \text{ \scriptsize hadrons}
	\end{array}
\endgroup
 \!\!}}^{}
a_i \sum_{l=0}^{\infty}\sum_{s=-s_i}^{s_i} \frac{|e_i|B}{2\pi}\int_{-\infty}^{\infty}\frac{dk_z}{2\pi} \ln\FB{1+a_ie^{-\beta\omega^i_{kls}}}
\label{eq.Pot.B}
\end{eqnarray}
where, the second term on the right hand side (RHS) corresponds to the contribution from the neutral hadrons which are not affected by the external magnetic field in the leading order, and $\Omega^B_\text{vac}$ is given by
\begin{eqnarray}
	\Omega^B_\text{vac} = -\frac{B^2}{2}
	- \sum_{i\in\SB{ \!\!\!
			\begingroup
			\renewcommand*{\arraystretch}{0.55}		
			\begin{array}{c}
				\text{\scriptsize charged} \\ \text{ \scriptsize hadrons}
			\end{array}
			\endgroup
			\!\!}}^{}
	a_i \sum_{l=0}^{\infty}\sum_{s=-s_i}^{s_i} \frac{|e_i|B}{2\pi}\int_{-\infty}^{\infty}\frac{dk_z}{2\pi} \omega^i_{kls} 
	\label{eq.Pot.B.Vac}
\end{eqnarray}
which is divergent and independent of temperature but depends explicitly on magnetic field. Similar to the $B=0$ case, the contribution of $\Omega^B_\text{vac}$ to the thermodynamic variables will be subtracted out while defining normalized quantities.

In the presence of external magnetic field, the pressure becomes anisotropic and have different values along the longitudinal and transverse direction with respect to the direction of the magnetic field~\cite{Strickland:2012vu}. The longitudinal pressure is obtained from $P_\parallel = -\Omega_B$ whereas the transverse pressure is $P_\perp = P_\parallel-MB$ where $M=\FB{\frac{\del P_\parallel}{\del B}}$ is the magnetization of the system. Thus, the explicit expression of transverse pressure comes out to be
\begin{eqnarray}
P_\perp = P_{\perp\text{vac}}^B +  P_{\!\!\!\!\!\begingroup
	\renewcommand*{\arraystretch}{0.55}		
	\begin{array}{c}
	\text{\scriptsize neutral} \\ \text{ \scriptsize hadrons}
	\end{array}
	\endgroup}
+ \sum_{i\in\SB{ \!\!\!
		\begingroup
		\renewcommand*{\arraystretch}{0.55}		
		\begin{array}{c}
		\text{\scriptsize charged} \\ \text{ \scriptsize hadrons}
		\end{array}
		\endgroup
		\!\!}}^{}
 \sum_{l=0}^{\infty}\sum_{s=-s_i}^{s_i} \frac{|e_i|B}{2\pi}\int_{-\infty}^{\infty}\frac{dk_z}{2\pi} 
 \frac{1}{2\omega^i_{kls}} \{2l+1-2s~\sign{e_i}\}|e_i|B \frac{1}{e^{\beta\omega^i_{kls}}+a_i}
 \label{p.perp}
\end{eqnarray}
where the first term on the RHS is the contribution to the transverse pressure from $\Omega^B_\text{vac}$, and second term on the RHS corresponds to the contribution of isotropic pressure from the neutral hadrons. All the other thermodynamic quantities of interest can be calculated from Eq.~\eqref{eq.Pot.B}. The energy density in this case is given by
\begin{eqnarray}
\varepsilon =\varepsilon_\text{vac}^B + \varepsilon_{\!\!\!\!\begingroup
	\renewcommand*{\arraystretch}{0.55}		
	\begin{array}{c}
	\text{\scriptsize neutral} \\ \text{ \scriptsize hadrons}
	\end{array}
	\endgroup} 
+ \sum_{i\in\SB{ \!\!\!
		\begingroup
		\renewcommand*{\arraystretch}{0.55}		
		\begin{array}{c}
		\text{\scriptsize charged} \\ \text{ \scriptsize hadrons}
		\end{array}
		\endgroup
		\!\!}}^{}
 \sum_{l=0}^{\infty}\sum_{s=-s_i}^{s_i} \frac{|e_i|B}{2\pi}\int_{-\infty}^{\infty}\frac{dk_z}{2\pi} 
 \frac{\omega^i_{kls}}{e^{\beta\omega^i_{kls}}+a_i}	
 \label{en}
\end{eqnarray} 
where the first term on the RHS is the contribution to the energy density from $\Omega^B_\text{vac}$, and the second term on the RHS again corresponds to the contribution to energy density from the neutral hadrons.

Till now, we have considered the system size to be infinite. To take into account the finite-volume effect in HRG thermodynamics, we employ the formalism of generalized Matsubara prescription as discussed in Ref.~\cite{Abreu:2019czp}. For this, we consider our system to be a cube of length $L$ so that, the spatial coordinates lie in the interval $x^i\in[0,L]$. As a consequence of the generalized Matsubara prescription~\cite{Bellac:2011kqa,Kapusta:2006pm,Mallik:2016anp,Khanna:2014qqa}, the momentum integral at $B=0$ will have to be replaced with sum over discrete Matsubara modes as
\begin{eqnarray}
\int\frac{d^3k}{(2\pi)^3}f(\vec{k}) \to \frac{1}{L^3} \sum_{n_x=-\infty}^{\infty}\sum_{n_y=-\infty}^{\infty}\sum_{n_z=-\infty}^{\infty} f(\vec{k}_{n_xn_yn_z})
\label{int2sum0}
\end{eqnarray}
where,
\begin{eqnarray}
\vec{k}_{n_xn_yn_z} = \frac{2\pi}{L} \TB{ (n_x + b)\hat{x} + (n_y + b)\hat{y} + (n_z + b)\hat{z} }
\label{eq.kn}
\end{eqnarray}
in which the parameter $b$ can be chosen appropriately to consider periodic or anti-periodic boundary condition in the compactified spatial coordinates. It is well known from the Kubo-Martin-Schwinger (KMS) relation~\cite{Bellac:2011kqa,Kapusta:2006pm,Mallik:2016anp} in the context of finite temperature field theory, that the imaginary time coordinate ($\tau=it$) becomes periodic (anti-periodic) for Bosonic (Fermionic) system. However, no restrictions are applied for the compactified spatial coordinates. Though, the Lattice QCD calculations~\cite{Klein:2017shl} generally employ the periodic boundary condition on the sptial coordinates of Fermions, yet other work on QCD effective models~\cite{Wang:2018qyq,Shi:2018swj,Gasser:1986vb} takes the identical boundary conditions in the temporal as well as spatial coordinates. Hence, following Ref.~\cite{Abreu:2019czp}, in this work we consider periodic boundary condition for the mesons and anti-periodic boundary condition for the baryons. Hence, in our calculation, we choose the parameter $b$ in Eq.~\eqref{eq.kn} as
\begin{eqnarray}
b = \begin{cases}
0 ~~~\text{for Mesons,} \\
1/2 ~~~\text{for Baryons.}
\end{cases}
\end{eqnarray}

For, non-zero external magnetic field, due to dimensional reduction, the transverse momentum is already Landau quantized. Thus, the following Matsubara prescription has been used for a magnetized HRG with finite system
\begin{eqnarray}
\int_{-\infty}^{\infty}\frac{dk_z}{2\pi}f(k_z) \to \frac{1}{L} \sum_{n_z=-\infty}^{\infty} f(k_{z;n_z})
\end{eqnarray}
where,
\begin{eqnarray}
k_{z;n_z} = \frac{2\pi}{L}(n_z+b).
\end{eqnarray}
\begin{figure}[h]
	\begin{center}
		\includegraphics[angle=0,scale=0.32]{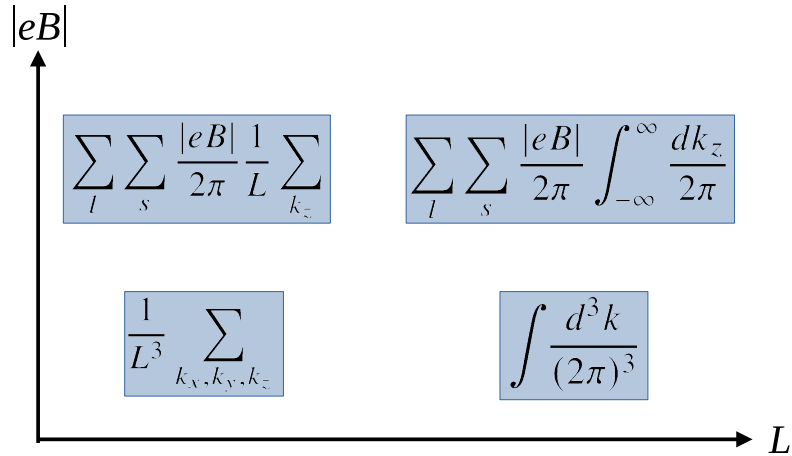}
	\end{center}
	\caption{A cartoon showing the modification of phase space at different regions in the $|eB|-L$ plane.}
	\label{fig.0}
\end{figure}

Few comments on the simultaneous consideration of finite size effect and external magnetic field on a general thermodynamic system of charged particles are in order here. 
At $B=0$, all the spatial coordinates $(x,y,z)$ appear as cyclic coordinates in the Equation of Motions (EOMs) for the fields describing particles with different spins. For example,
\begin{eqnarray}
\text{Klein-Gordon equation:}~~ (\del_\mu\del^\mu+m^2)\phi(x)=0 ~~\text{for spin-0 mesons, } \nn \\
\text{Dirac equation: } ~~ (i\gamma^\mu\del_\mu-m)\psi(x)=0 ~~\text{for spin-1/2 baryons, } \nn \\
\text{Maxwell-Proca equation: } ~~ (\del_\mu\del^\mu+m^2)A^\nu(x)=0 ~~ \text{for spin-1 mesons,} \nn
\end{eqnarray}
etc. in which all the $(x,y,z)$ are cyclic. For this, one can construct well defined momentum eigen-states $\ket{k_x,k_y,k_z}$ which in-turn enables us to easily quantize all the momentum components as per Eq.~\eqref{eq.kn} while limiting the system size to a cubical box of length $L$.

The scenario becomes complicated in presence of external magnetic field. The EOMs will now contain the covariant derivative $D^\mu = \del^\mu + iqA^\mu_\text{ext}$, where $q$ is the electronic charge of the particle and $A^\mu_\text{ext}$ is the four potential that generates the external magnetic field. As we have considered constant $\vec{B}=B\hat{z}$ in the positive $\hat{z}$-direction, it can be achieved by different choice of gauges. If we chose the Landau gauge in which the four-potential is $A^\mu_\text{ext} = (0,0,Bx,0)$, then the EOMs will explicitly contain the spatial coordinate $x$ whereas the other two coordinates $(y,z)$ will remain cyclic. In that case, the momentum eigen-states will be of the form $\ket{l,s,k_y,k_z}$ though the energy eigenvalue is independent of $k_y$~\cite{Andersen:2014xxa}. Considering the system with finite volume $V=L^3$, it has also been shown in Ref.~\cite{Andersen:2014xxa} that, the characteristic size of a Landau level is $1/\sqrt{|qB|}$ and the degeneracy associated with the quantum number $k_y$ is $N=\frac{|qB|}{2\pi}L^2$ so that the the phase space integral is modified as per~\cite{Andersen:2014xxa,Abreu:2019czp,Abreu:2013nca}
\begin{eqnarray}
	\int\frac{d^3k}{(2\pi)^3} \to \frac{1}{V} \sum_{k_x,k_y,k_z}^{}  \to \frac{1}{V} \frac{|qB|}{2\pi} L^2 \sum_{s=-s_i}^{s_i} \sum_{l=0}^{\infty} \sum_{k_z}^{} 
	 =  \frac{|qB|}{2\pi} \sum_{s=-s_i}^{s_i} \sum_{l=0}^{\infty} \frac{1}{L} \sum_{k_z}^{}.
	\label{eq.phase.space.2}
\end{eqnarray} 
If we have chosen a different gauge for example, $A^\mu_\text{ext} = (0,-By,0,0)$, then the momentum eigen-states would have been $\ket{l,s,k_x,k_z}$; however, the above result would still hold and is  independent of the choice of gauge. A cartoon describing the modification of phase space at different regions in the $|eB|-L$ plane has been shown in Fig.~\ref{fig.0}.

%
%

\section{NUMERICAL RESULTS \& DISCUSSIONS}\label{sec.results}
\begin{figure}[h]
	\begin{center}
		\includegraphics[angle=-90,scale=0.35]{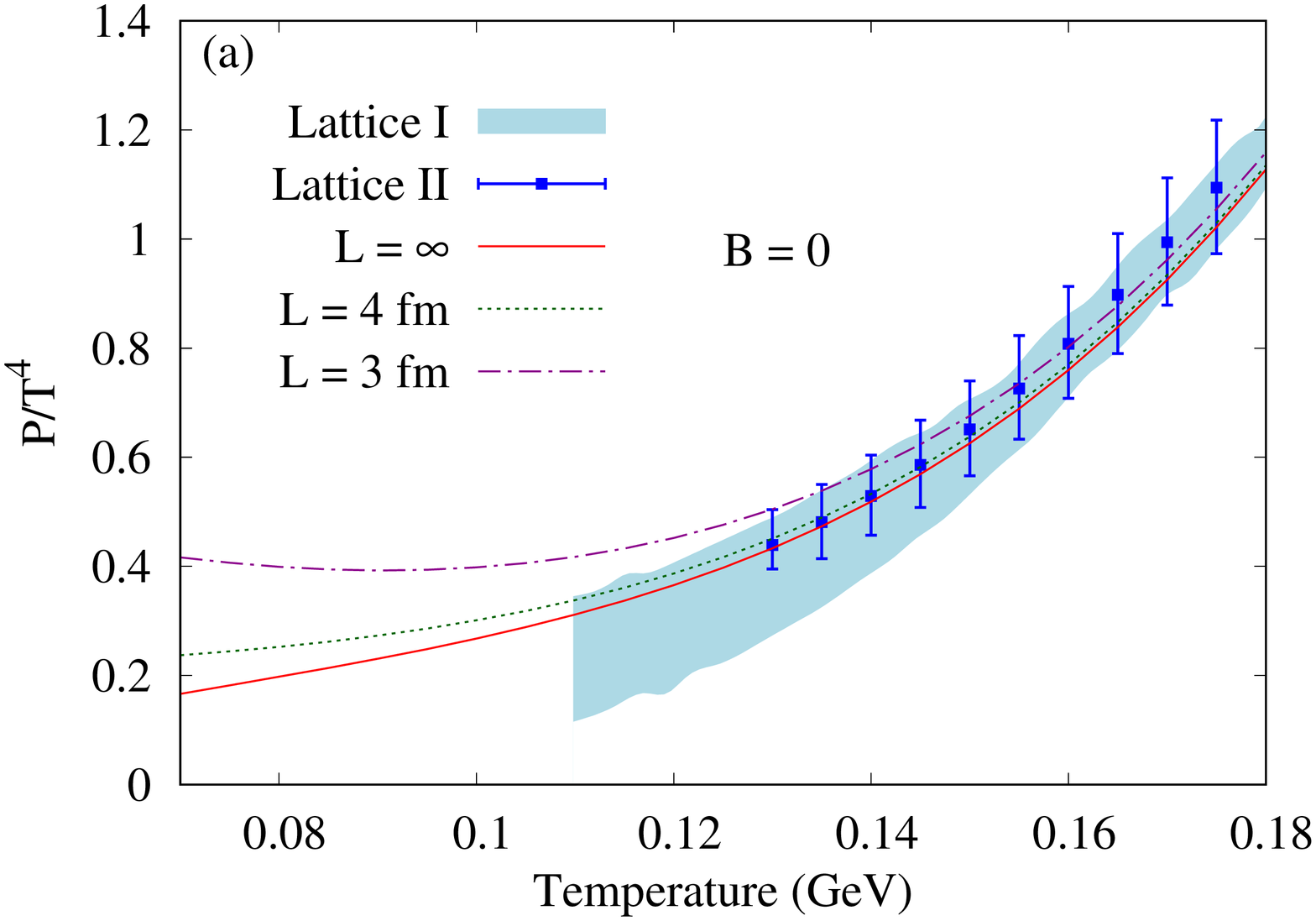} \includegraphics[angle=-90,scale=0.35]{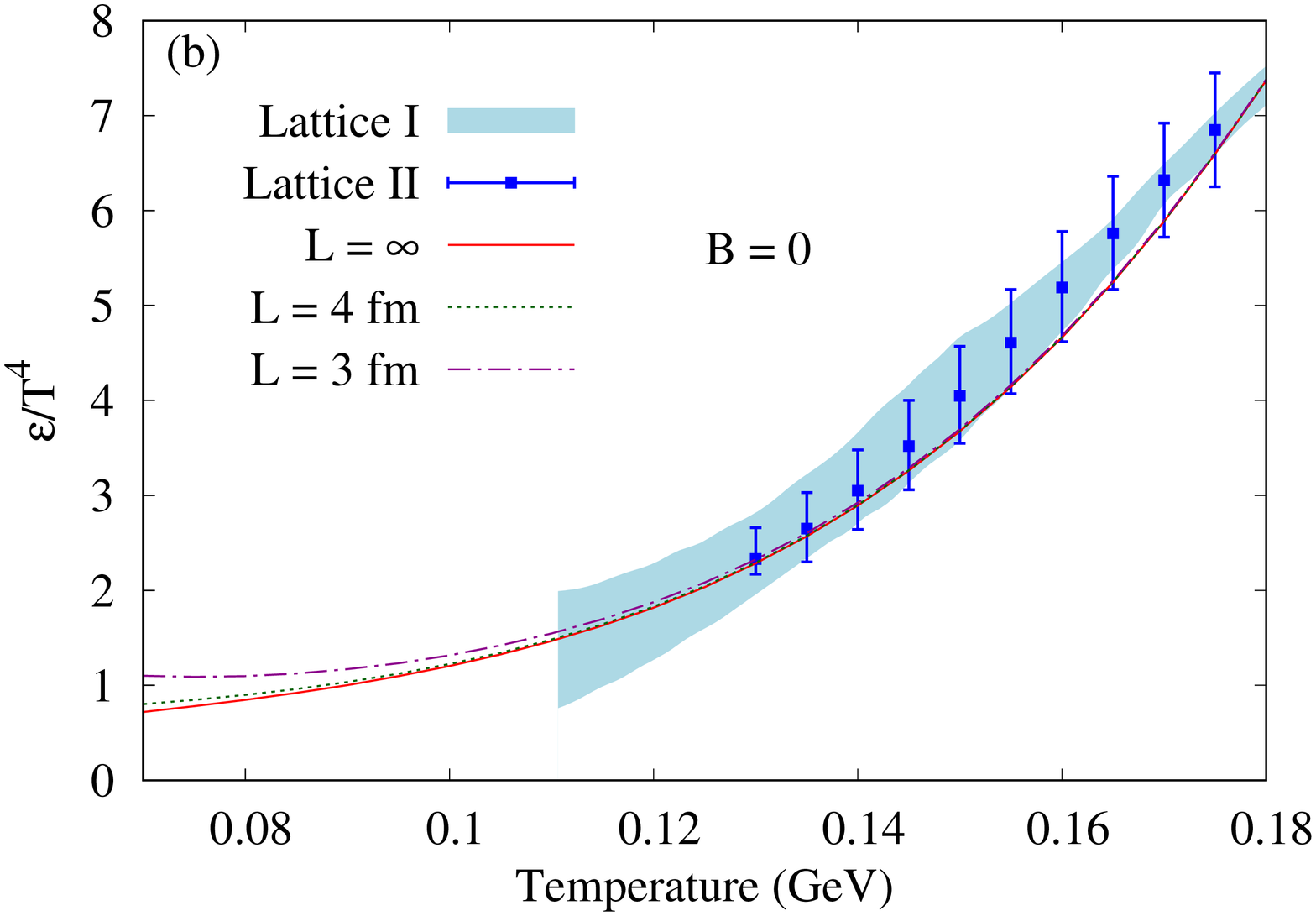} 
	\end{center}
	\caption{The variation of normalized (a) $P/T^4$ and (b) $\varepsilon/T^4$ as a function of temperature ($T$) for different values of system size ($L$) at $B=0$. For comparison, the results of Lattice QCD calculations from Ref.~\cite{Borsanyi:2013bia} and Ref.~\cite{Bazavov:2014pvz} are also shown as Lattice I and Lattice II respectively.}
	\label{fig.1}
\end{figure}

We begin this section by showing the variations of dimensionless scaled pressure $\frac{P}{T^4}$ and energy density $\frac{\varepsilon}{T^4}$ as a function of temperature at zero magnetic field for different values of system sizes in Figs.~\ref{fig.1}(a) and (b). For the numerical calculation, we have considered all the hadrons upto mass 2.6 GeV from Ref.~\cite{ParticleDataGroup:2018ovx}.
As already mentioned in Sec.~\ref{sec.formalism}, all the thermodynamic quantities contain a divergent vacuum term which needs to be subtracted out. Thus, at $B=0$, a given thermodynamic quantity $\Xi \in \{\varepsilon,P,s,\cdots\}$, is normalized/scaled as per
\begin{eqnarray}
\Xi_N(T) = \Xi(T) - \Xi(T=0). 
\label{eq.normalization.0}
\end{eqnarray} 
From now on, we will suppress the subscript `$N$' and all the results presented in this section will be of normalized thermodynamic quantities. Eq.~\eqref{eq.normalization.0} ensures that $\Xi_N$ vanishes at $T=0$. First of all, we notice that, both the pressure and energy density increase monotonically with the increase in temperature. This is due to the enhancement of thermal phase space of hadrons owing to the increase in their average thermal energy.  It is evident from Figs.~\ref{fig.1} (a) and (b) that  the results from HRG model for infinite size system ($L = \infty$) are in excellent agreement with the corresponding LQCD estimates of Refs.~\cite{Borsanyi:2013bia} and \cite{Bazavov:2014pvz} which are denoted as Lattice I and Lattice II respectively in Figs.~\ref{fig.1}(a) and (b). This observation supports the previously mentioned key feature of the HRG model that the HRG-thermodynamics is quite close to the LQCD especially in the low temperature regions. 
\begin{figure}[h]
	\begin{center}
		\includegraphics[angle=-90,scale=0.35]{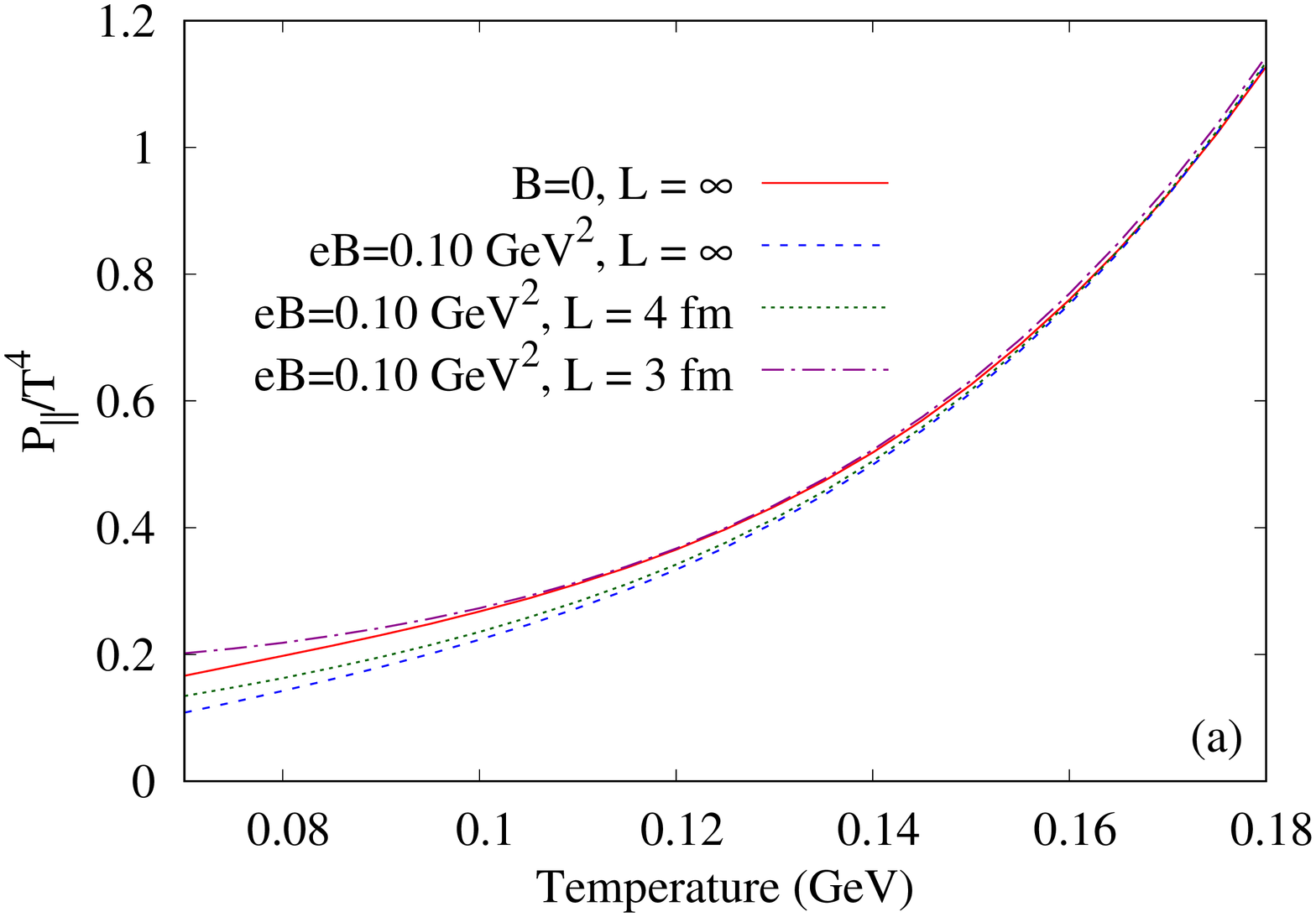} \includegraphics[angle=-90,scale=0.35]{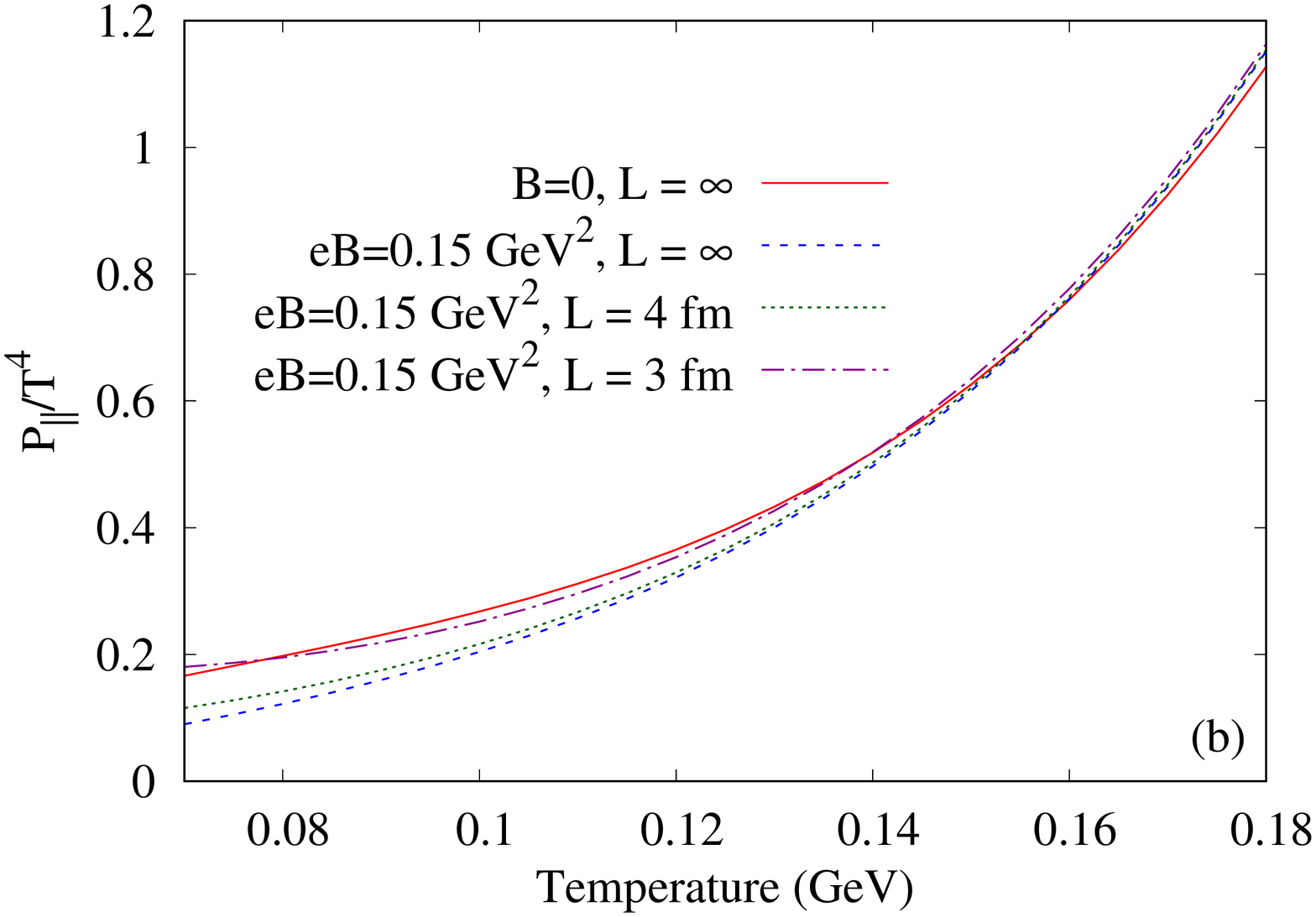} 
		\includegraphics[angle=-90,scale=0.35]{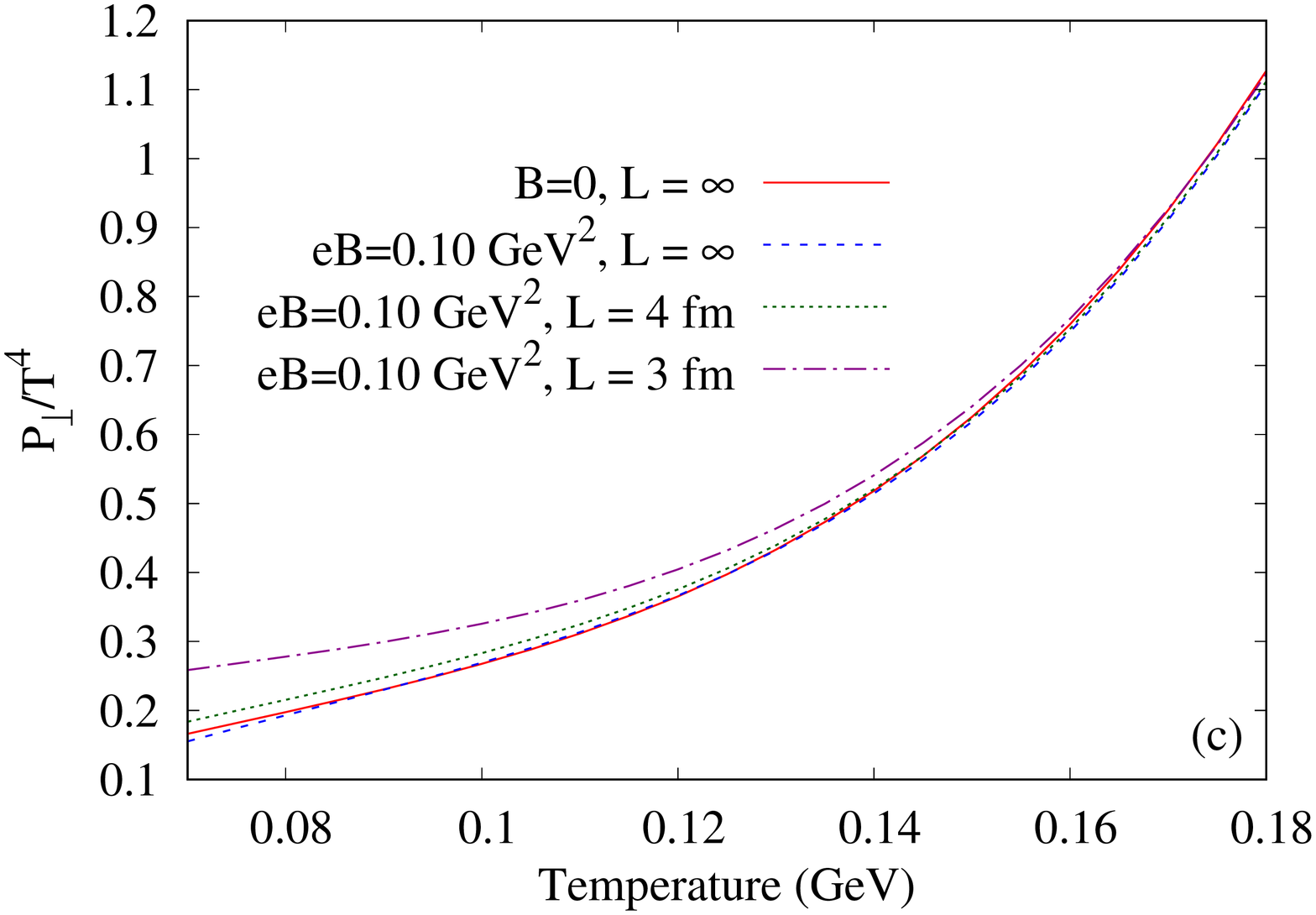} \includegraphics[angle=-90,scale=0.35]{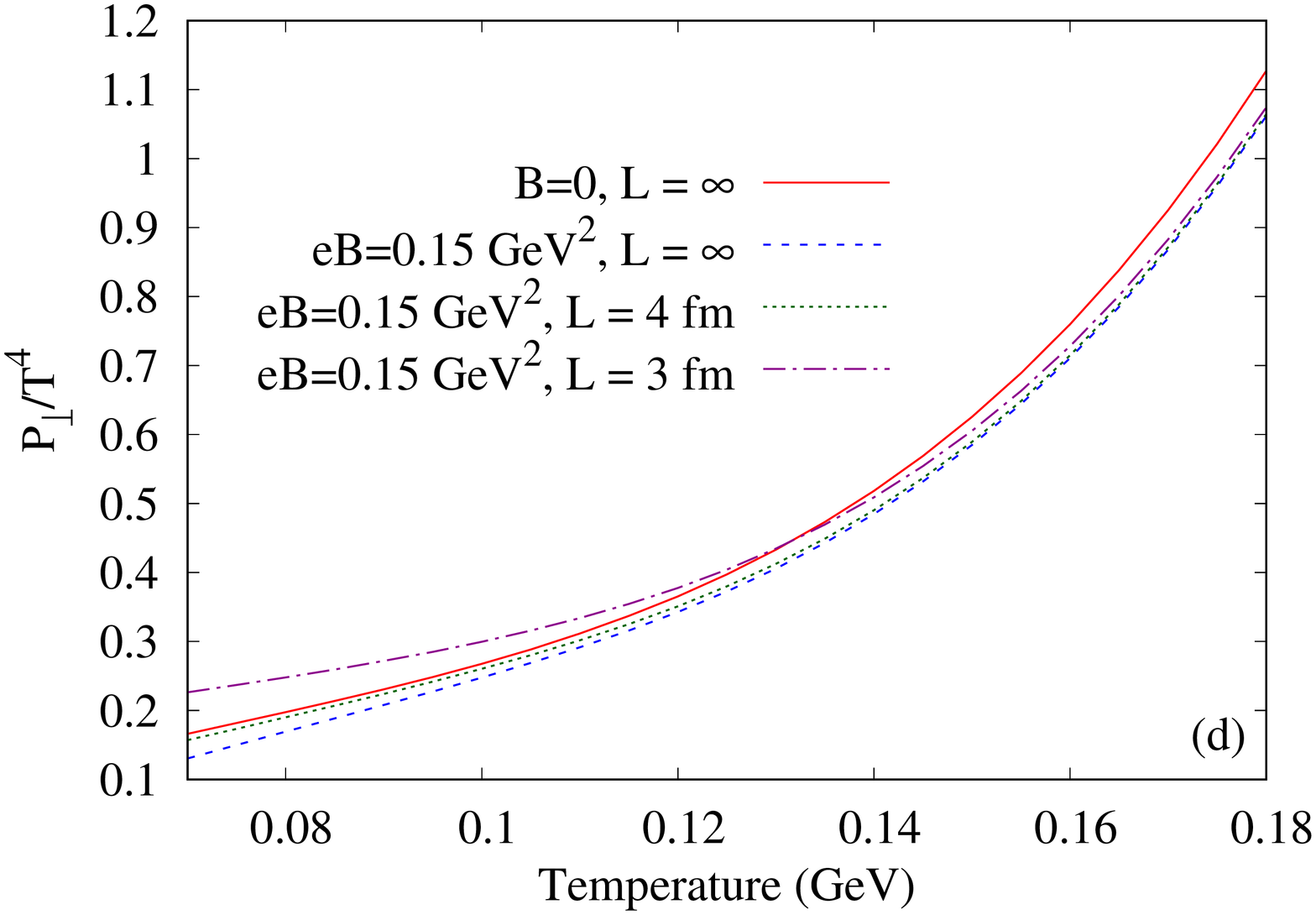} 
		\includegraphics[angle=-90,scale=0.35]{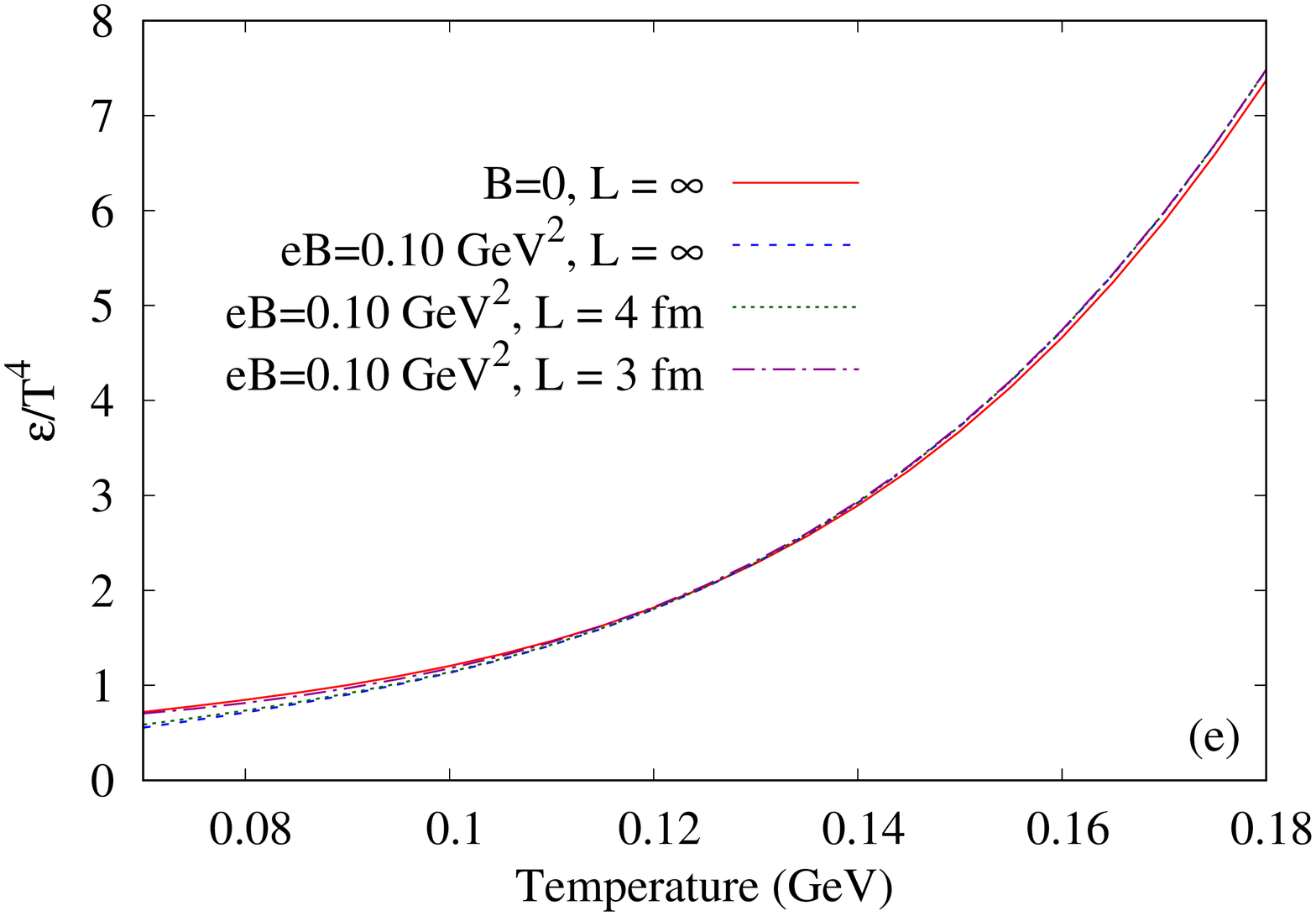} \includegraphics[angle=-90,scale=0.35]{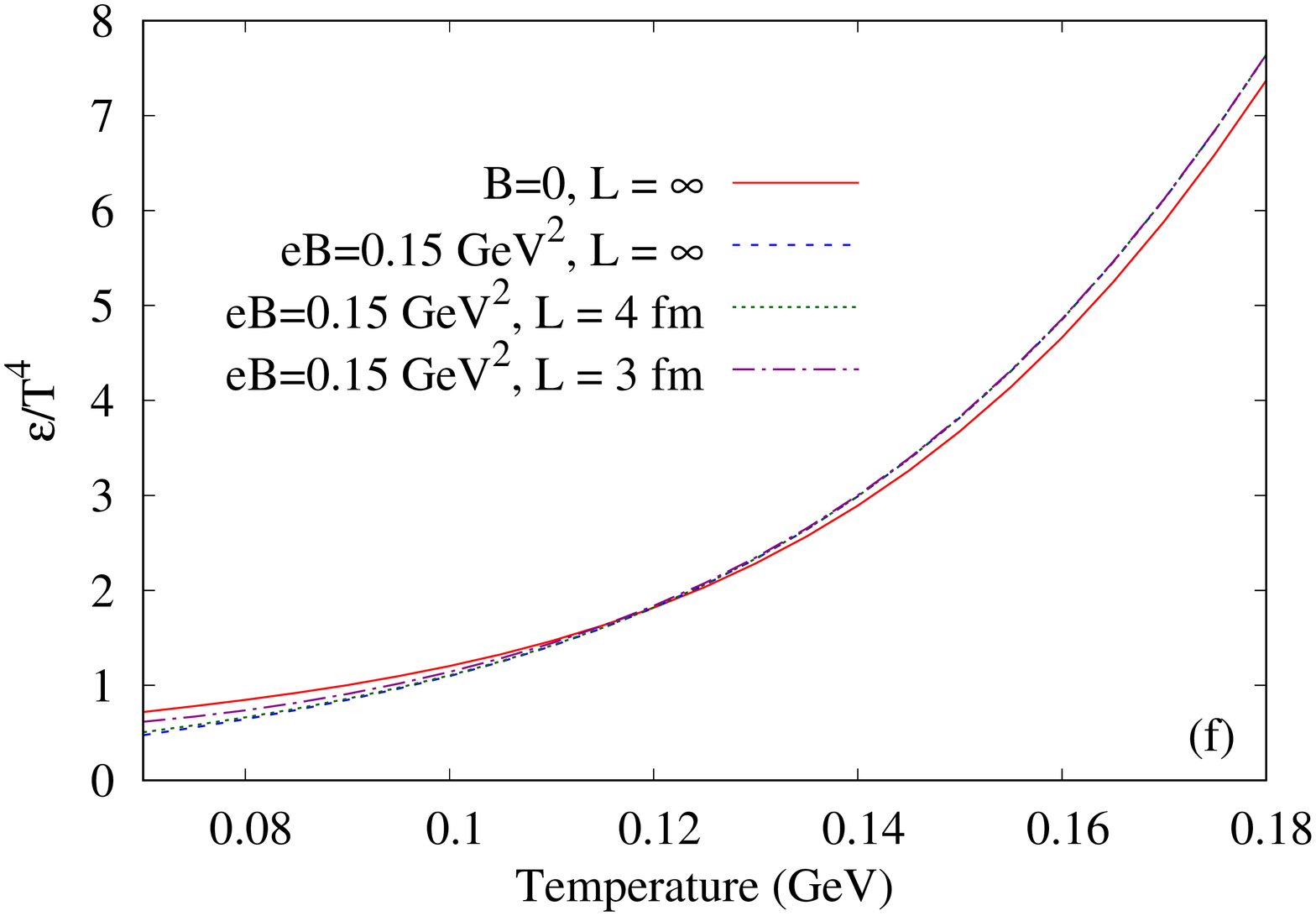}  
	\end{center}
	\caption{The variation of normalized (a) and (b) $P_\parallel/T^4$, (c) and (d) $P_\perp/T^4$, and, (e) and (f) $\varepsilon/T^4$ as a function of temperature ($T$) for different values of system size ($L$) and magnetic field $B$.}
	\label{fig.2}
\end{figure}

In Figs.~\ref{fig.1}(a) and (b), we have also depicted the temperature dependence of scaled pressure and energy density for system with finite size. It can be observed that, the overall magnitude of $ P/T^4 $ and $ \varepsilon/T^4 $ increases substantially as the system size decreases. This behaviour can be understood from the following arguments. The expressions of $P$ and $\varepsilon$ at $L=\infty$ in Eqs.~\eqref{eq.Pot.0} and \eqref{e0} contain integrations over momentum. The corresponding integrands are monotonically decreasing functions of the (magnitude of) momentum since the high momentum modes are Boltzmann suppressed. When we restrict the system size to be finite, we essentially discretize the momentum according to Eq.~\eqref{eq.kn} and the integration is replaced by a summation as per Eq.~\eqref{int2sum0}. From the physical interpretation of definite integral (in terms of area under the curve), it is well known that, left Riemann sum (in Eq.~\eqref{int2sum0}) overestimates the integral (the area under the curve) for a monotonically decreasing function. This explains the enhancement of $P$ and $\varepsilon$ for a finite sized system as compared to the $L=\infty$ case.

Moreover a close examination of Figs.~\ref{fig.1}(a) and (b) reveals that, the effect of the finiteness of the system size on the scaled pressure and energy density are more prominent in the low temperature region. This observation can also be explained in the similar manner by using the area under the curve interpretation of integration. In Eq.~\eqref{int2sum0}, the amount of overestimation of the left Riemann sum over the integral for a monotonic but rapidly decreasing function will be larger compared to the slowly decreasing one. Now, for low values of temperature, the Boltzmann suppression is more effective. As a result, the integrands in the expressions of $ P $ and $ \varepsilon $ will composed of a rapidly decreasing function of momentum. This leads to a considerable change in the magnitude of the scaled pressure and energy density in the low temperature region and thus explains the noticeable change in their magnitudes at small temperature values. On the other hand, at high values of temperature, Boltzmann suppression is weak. Hence, the consideration of finite size system does not bring substantial change in the $ T $-dependence of  $ P/T^4 $ and $ \varepsilon/T^4 $ in the high temperature region. 

Next part of this section is devoted to show the effects of the presence of finite background magnetic field on the thermodynamics of HRG model for both finite and infinite size systems. 
In the numerical calculations, the magnetic field effects are taken only for charged mesons with spin $0$ and $1$, and for baryons with spins $1/2$ and $3/2$. Hadrons with higher spins are not affected considerably by the external magnetic field (of expected magnitude in HIC) due to their large masses.
At $B\ne0$, the normalization of thermodynamic quantities are done as per 
\begin{eqnarray}
	\Xi_N(T,B) = \Xi(T,B) - \Xi(T=0,B),
	\label{eq.normalization.1}
\end{eqnarray} 
and as before we will suppress the subscript `$N$' while discussing numerical results. We have already mentioned in Sec.~\ref{sec.formalism} that, the presence of the background magnetic field results in different values of pressure in longitudinal and transverse direction with respect to the magnetic field. In Figs.~\ref{fig.2}~(a) and (b) we have demonstrated the variation of the dimensionless scaled longitudinal pressure  $\frac{P_\parallel}{T^4}$ as a function of temperature considering different system sizes for $ eB =0.10 $ and $ 0.15 $ GeV$ ^2 $ respectively.  The different values of background field chosen here are relevant in the context of  HIC phenomenology. For comparison we have also plotted the $ T $-dependence of the total scaled pressure at vanishing magnetic field for infinite size system.  Form Figs.~\ref{fig.2}~(a) and (b), as we increase  the value of the background field, a non-monotonic variation of the scaled longitudinal pressure for different temperature value is evident. It can be seen that, in the low temperature region, the magnitude of $\frac{P_\parallel}{T^4}$ decreases with increase in $ eB $. However, an opposite phenomena is observed in the high temperature region. This observation is consistent with Ref.~\cite{Tawfik:2016lih}. The reason of the non-monotonicity is the presence of $\frac{eB}{T^2}$ in the exponential in Eqs.~\eqref{eq.Pot.B}-\eqref{en} which is mainly controlling the $B$-dependence of the thermodynamic quantities at different temperatures. Furthermore, one can notice that, with the decrease in system size $L$, the scaled longitudinal pressure at non-zero external magnetic field increases as can be seen by comparing blue (dash), green (dot) and magenta (dash-dot) curves in Fig.~\ref{fig.2}. Similar to the zero magnetic field case, the finite size effect on magnetized HRG is more prominent in the low temperature region. The last two features are common in all the thermodynamic quantities at non-zero external magnetic field.  Now in Figs.~\ref{fig.2}~(c) and (d), the $ T $-dependence of the scaled transverse pressure $\frac{P_\perp}{T^4}$ is shown for different values of magnetic field and system sizes. From both the figures, it is evident that, unlike the longitudinal pressure, the transverse pressure shows almost monotonic behaviour with magnetic field which is understandable owing to the presence of an extra factor of $eB$ in the numerator of Eq.~\eqref{p.perp} giving additional $B$-dependence. The same non-monotonic behaviour can be seen again in the variation of scaled energy density   $\frac{\varepsilon}{T^4}$  as a function of temperature as can be seen from Figs.~(e) and (f). It should be noted that, the energy density is least affected by the finiteness of the system size compared to logitudinal and transverse pressure.

Few comments on the normalization of the thermodynamic quantities in presence of external magnetic field are in order here. Following Refs.~\cite{Avancini:2020xqe,Chaudhuri:2020lga}, we have used Eq.~\eqref{eq.normalization.1} which ensures that $\Xi_N(T,B)$ vanishes at $T=0$ and is clearly visible in Fig.~\ref{fig.2}. However, there exists a much elegant normalization prescription that has been used in Refs.~\cite{Endrodi:2013cs,Kadam:2019rzo}: 
\begin{eqnarray}
\Xi_R(T,B) = \Xi(T,B) - \Xi(T=0,B=0)
\label{eq.normalization.2}
\end{eqnarray} 
where, $\Xi_R(T,B)$ may be called ``regularized'' thermodynamic quantity. The calculation of $\Xi_R(T,B)$ is bit involved and requires the dimensional regularization of $\Xi(T,B)$ in order to subtract the pure vacuum part $ \Xi(T=0,B=0)$. Moreover, Eq.~\eqref{eq.normalization.2} implies that at $T=0$, $\Xi_R(T=0,B)$ will be non-zero and will depend on the external magnetic field~\cite{Endrodi:2013cs,Kadam:2019rzo} in contrast to the behaviour of $\Xi_N(T=0,B)$ with is by definition zero at $T=0$ from Eq.~\eqref{eq.normalization.1}.

\begin{figure}[h]
	\begin{center}
		\includegraphics[angle=-90,scale=0.35]{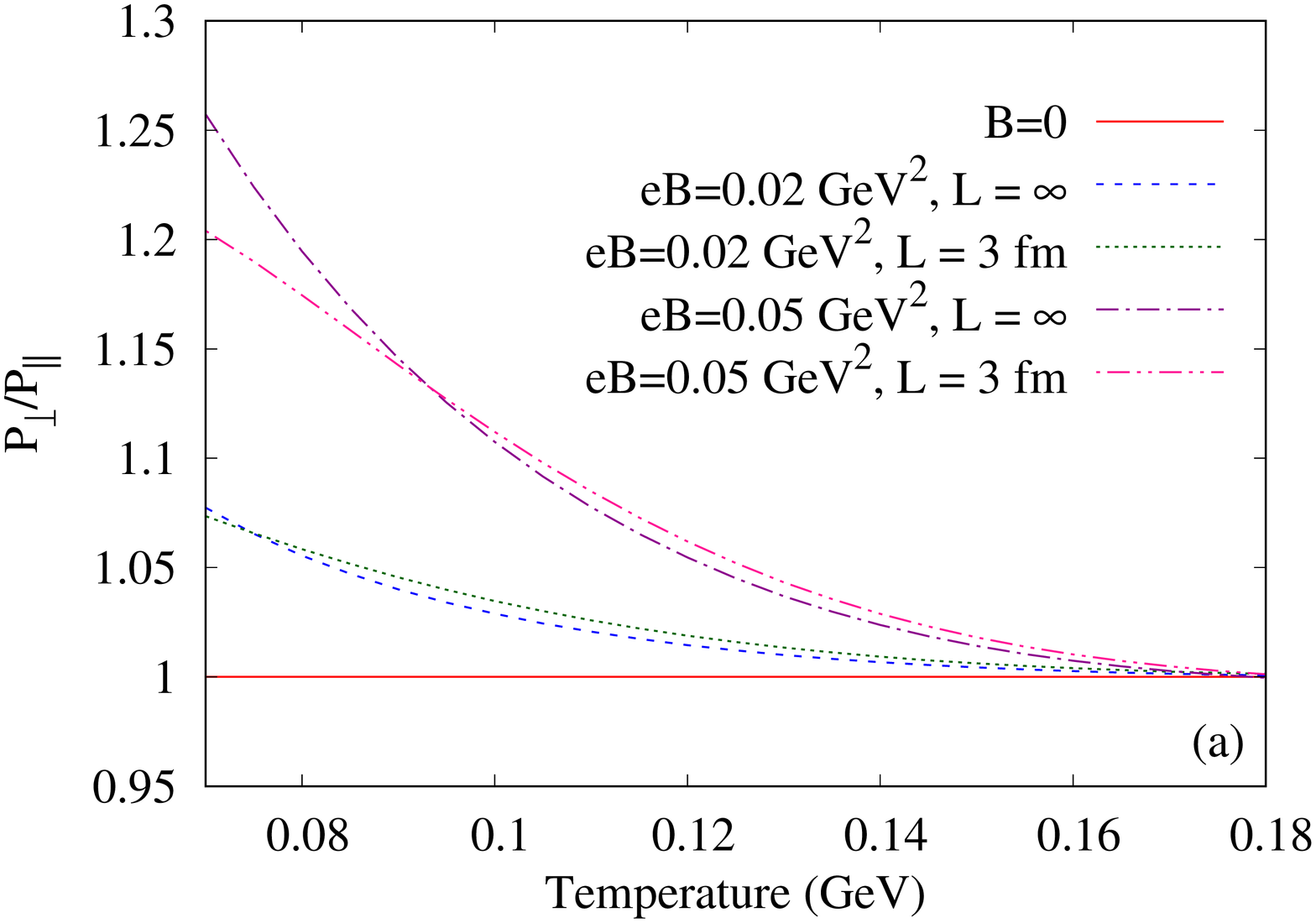} 
			\includegraphics[angle=-90,scale=0.35]{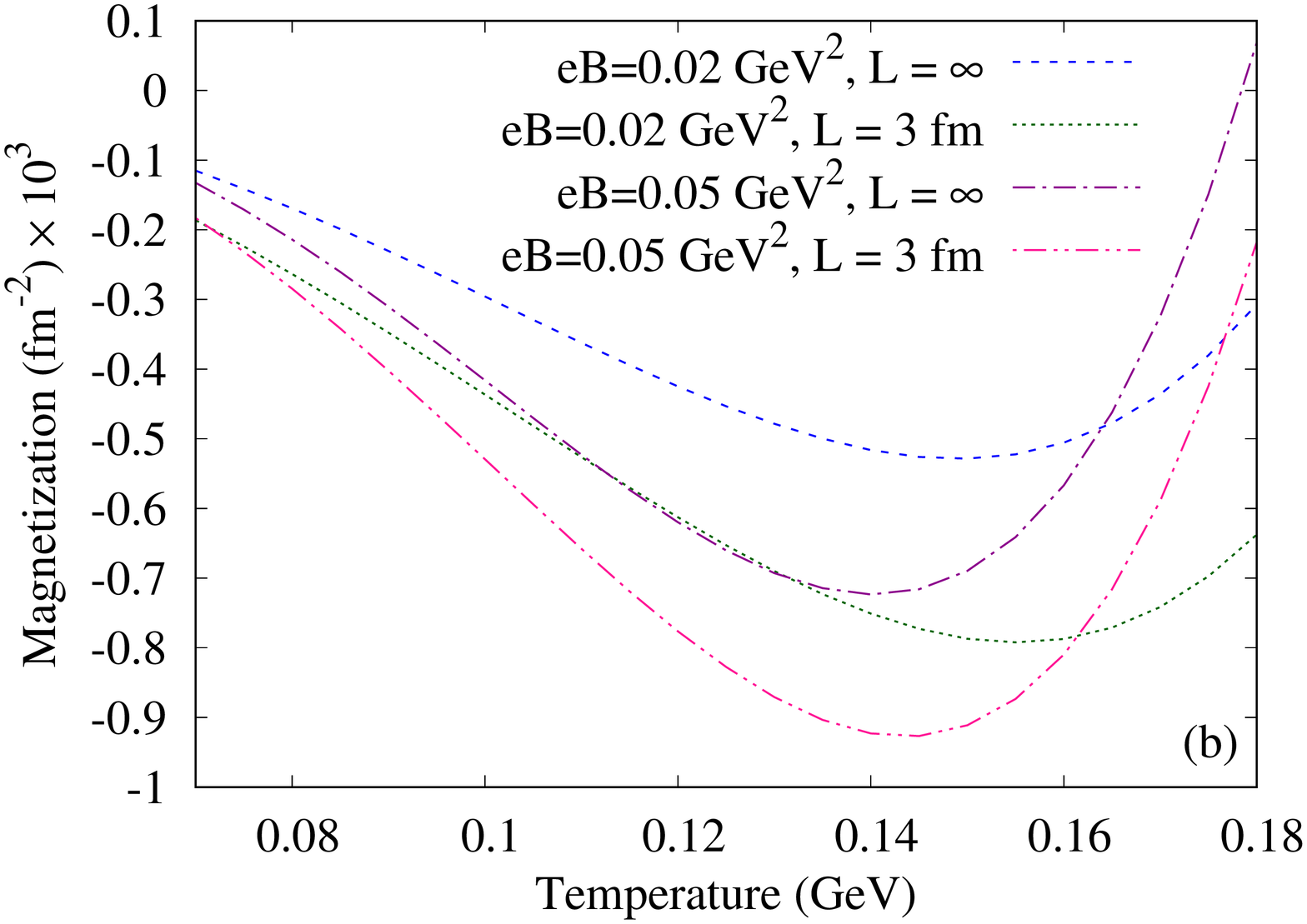} 
	\end{center}
	\caption{The variation of (a) the ratio $P_\perp/P_\parallel$ and (b) magnetization as a function of temperature ($T$) for different values of system size ($L$) and magnetic field $B$. }
	\label{fig.3}
\end{figure}

 Next in Fig.~\ref{fig.3}(a), we have shown the ratio $\frac{P_\perp}{P_\parallel}$ as a function of temperature for different values of magnetic field and system size. This ratio may be considered as a measure of the anisotropy introduced by the magnetic field  in the system (since at $B=0$, this ratio is unity). From Fig.~\ref{fig.3}(a), it is evident that, temperature and external magnetic field bring in opposite affects in this ratio. At first, we notice that, with the increase in magnetic field, independent of  system size, the magnitude of the ratio $\frac{P_\perp}{P_\parallel}$  increases  implying enhancement of the anisotropy in the system. On the contrary, an opposite behaviour is found when we increase the temperature of the system. This is due to the fact that, magnetic field brings anisotropy (orderness/alignment) in the medium while the temperature or thermal fluctuation tries to diminish it owing to bring disorder and randomness in the system. At high enough temperature the ratio should tend to unity. Also, the effect of system size on the anisotropy measure is found to be different at different temperature. We have found that, smaller systems are more anisotropic (isotropic) in the high (low) temperature region.

Finally, in Fig.~\ref{fig.3}(b), we have plotted the normalized magnetization $M = \FB{\frac{\del P_\parallel}{\del B}}$ as a function of temperature for different values of magnetic field and system size. Qualitatively, for all the cases, we notice that, $M$ is negative in the low temperature region; moreover, with the increase in temperature, $M$ first decreases, attains a local minima and then increases. Also, for a particular system size, at lower values of temperature, $M$ decreases with the increase in $B$ whereas an opposite effect is seen for the higher temperature values owing to crossing of the curves. This is consistent with the observation made in Fig.~\ref{fig.2}(a) and (b) in which similar type of crossing of the curves have been noticed. Such variation of $M$ with $B$ implies that, the susceptibility of the magnetized HRG defined via $\chi=\FB{\frac{\del M}{\del B}}$ will be negative ($\chi<0$) in the low temperature region exhibiting a diamagnetic behaviour whereas $\chi>0$ in the high temperature region indicating a paramagnetic behaviour of the medium. The diamagnetic nature at lower temperature agrees with the LQCD prediction~\cite{Bali:2014kia}. Interestingly, the observed paramagnetism at high temperature is also consistent with the LQCD calculations~\cite{Bali:2014kia} though the use of the HRG model at such high-$T$ region is not well justified. The consideration of finite system size further strengthen the diamagnetic nature of the medium.

Few comments on the consideration of the fermions while calculating the thermodynamic quantities in presence of magnetic field are in order here. In Ref.~\cite{Endrodi:2013cs}, it has been argued that, the contribution to the pressure for spin-$ 3/2 $ fermions can be negative for nonzero magnetic fields. However, we found that, for spin-$ 3/2 $ fermions, the longitudinal pressure ($ P_\parallel $) remains positive for any values of magnetic field whereas the transverse pressure ($ P_\perp $) becomes negative for high values of $ B $. Moreover, the negativity of fermionic transverse pressure has widely been observed in the literature even for spin-$ 1/2 $ fermions for example in Refs.~\cite{Strickland:2012vu,Menezes:2015fla,Menezes:2015flaE,Huang:2009ue,Sinha:2013dfa,Ferrer:2015wca}.


\section{SUMMARY \& CONCLUSION}\label{sec.summary}
In summary, we have calculated the thermodynamic quantities (energy density, longitudinal and transverse pressure, pressure anisotropy, and, magnetization) of a non-interacting ideal HRG of finite volume in the presence of an external magnetic field. The dispersion relation of the charged handrons are modified due to the external magnetic field owing to the Landau quantization. To take into account the finite volume effect, we have employed the generalized Matsubara prescription in which a periodic (anti-periodic) boundary conditions is considered for the mesons (baryons). Due to boundary effects, the continuous momentum of the hadrons became discrete; so that the integrals over momentum became sums over discrete Matsubara modes.

We find that, energy density, longitudinal and transverse pressure increases with the decrease in system size. The finite size effects are found to be more in the low temperature region. For the ranges of magnetic field considered in this work: $0\le eB \le 0.15$ GeV$^2$ which is relevant for HIC experiments, we find that, with the increase in magnetic field, the energy density and longitudinal pressure decreases (increases) for the low (high) temperature region owing to a non-monotonic behaviour. Where as, the transverse pressure is found to decrease monotonically with the increase in external magnetic field. The ratio of the transverse to longitudinal pressure $\frac{P_\perp}{P_\parallel}$, which is a measure of anisotropy in the system, is found to increase (decrease) with the increase in magnetic field (temperature). This behaviour is consistent with the physical fact that, the magnetic field brings anisotropy (orderness/alignment) in the medium while the temperature or thermal fluctuation tries to diminish it owing to bring disorder, randomness and isotropy in the system. It has been found that, small magnetized system of hadrons is more anisotropic (isotropic) in the high (low) temperature region. Finally, the magnetization of the medium decreases with the decrease in system size whereas it shows non-monotonic behaviors with the variation of temperature and magnetic field owing to a diamagnetic (paramgetic) behavior in the low (high) temperature region. The finite size effect is seen to strengthen the diamagnetic nature of the medium.

\setstretch{1.0}
\section*{Acknowledgments}
The authors acknowledge Prof. Sourav Sarkar for the immense encouragement, support and useful discussions. We would like to thank the KANAAD computing facility of VECC, Kolkata. DA and SG are funded by the Department of Higher Education, Government of West Bengal, India. NC is funded by the Department of Atomic Energy (DAE), Government of India.

\bibliographystyle{apsrev4-1}
\bibliography{HRG-BL}

\end{document}